\documentclass[prl,nobibnotes,floatfix,superscriptaddress,twocolumn,showpacs]{revtex4-1}
\usepackage{amsfonts,amsmath,graphicx,epsfig,latexsym}
\usepackage{subfigure}
\usepackage{graphicx}
\usepackage{color}
\usepackage{natbib}
\usepackage{multirow}
\usepackage{latexsym,amssymb}
\usepackage{amsmath}

\begin{document}
\title{ Protracted Kondo coherence with dilute carrier density in Cerium based nickel pnictides}
\author{Peng Zhang}
\email{zpantz@mail.xjtu.edu.cn}
\affiliation{Department of Physics, School of Science, Xi'an Jiaotong University, Xi'an, 710049, Shaanxi, China}
\affiliation{MOE Key Laboratory for Nonequilibrium Synthesis and Modulation of Condensed Matter, School of Science, Xi'an Jiaotong University, Xi'an, 710049, Shaanxi, China}
\author{Bo Liu}
\affiliation{Department of Physics, School of Science, Xi'an Jiaotong University, Xi'an, 710049, Shaanxi, China}
\affiliation{Shaanxi Province Key Laboratory of Quantum Information and Quantum Optoelectronic Devices,
Xi'an Jiaotong University, Xi'an 710049, Shaanxi, China}
\author{Shengli Zhang}
\affiliation{Department of Physics, School of Science, Xi'an Jiaotong University, Xi'an, 710049, Shaanxi, China}
\affiliation{MOE Key Laboratory for Nonequilibrium Synthesis and Modulation of Condensed Matter, School of Science, Xi'an Jiaotong University, Xi'an, 710049, Shaanxi, China}

\author{K. Haule}
\affiliation{Department of Physics, Rutgers University, Piscataway, New Jersey 08854, USA}

\author{Jianhui Dai}
\email{daijh@hznu.edu.cn}
\affiliation{Department of Physics, Hangzhou Normal University, Hangzhou 310036, China}

\begin{abstract}
{\footnotesize

Nozi$\grave{e}$res' exhaustion theory argues the temperature for
coherently screening of all local moments in Kondo lattice could be much lower than the temperature of single moment screening with insufficient number of conduction electrons. Recent experiment [Luo et al, PNAS, 112,13520 (2015)] indicates the cerium based nickel pnictides $CeNi_{2-\delta}As_2 (\delta\approx0.28)$ with low carrier density is an ideal material to exam such protracted Kondo screening. Using the density functional theory and
dynamical mean-field theory, we calculated the respective electronic structures of paramagnetic $CeNi_2As_2$/$CeNi_2P_2$. In contrast to structurally analogous layered iron pnictides, the electronic structures of the present systems show strong three-dimensionality with substantially small contributions of Ni-3d electrons to the carrier density. Moreover, we find significant Kondo resonance peaks in the compressed $CeNi_2As_2$ and $CeNi_2P_2$ at low temperatures, accompanied by topological changes of the Fermi surfaces. We also find similar quantum phase transition in $CeNi_2As_2$ driven by chemical pressure via the isovalence As$\rightarrow$P substitution. }

\end{abstract}


\maketitle
\newpage

Recently extensive interests have been focused on the heavy fermion materials with dilute carrier density due to their exotic behaviors, for example the topologically nontrivial electronic states \cite{PhysRevLett.110.096401,PhysRevX.7.011027, PhysRevLett.118.246601}, the Kondo semimetals \cite{Rai2019, Lv2019, FengXY} and the quantum  phase transitions due to protracted Kondo screening \cite{Luo2012, Luo2014, Luo2015}. The properties of heavy fermion materials depend on two key factors, one is the electronic correlations between the localized f-electrons and another is the hybridization among the f-electrons and the conduction electrons \cite{Steglich1991, Hess1993, Hewson1993}. When hybridization is weak, the Ruderman-Kittel-Kasuya-Yosida (RKKY) interaction mediated by the hybridization leads to magnetic ordering of the localized f-electrons \cite{Ruderman1954,Kasuya1956,Yosida1957}. In the strong hybridization limit, the Kondo coupling forces the screening of local f-moments by conduction electrons producing a Fermi liquid at low temperature. However, the Kondo screening of localized f-electrons at the dilute conduction electron limit is less understood. According to Nozi$\grave{e}$res' argument \cite{Nozieres1985, Nozieres1998}, only the conduction electrons within the Kondo energy scale around the Fermi level can participate in the Kondo screening. When the charge density of conduction electrons is low, the number of available conduction electrons for Kondo screening can be smaller than the total number of localized f-electrons in the lattice. The full Kondo screening of all localized f-electrons, if happens, must be coherent. The corresponding energy scale for the coherent Kondo screening with insufficient conduction electrons is much lower than that of the single-impurity Kondo screening, being named the protracted Kondo screening temperature \cite{Sarrao1999,Lawrence2001}.

Previous theoretical investigations on Nozi$\grave{e}$res' argument are largely limited to simple models like the single-impurity Anderson model, the periodic Anderson lattice model, and the Kondo lattice model \cite{Tahvildar-Zadeh1997,Tahvildar-Zadeh1998, PhysRevB.60.10782, PhysRevB.61.12799, Vidhyadhiraja2000, Burdin2000}. Two energy scales are introduced in these calculations: the single-impurity Kondo temperature, $T_K$, that indicates the screening of a localized f-electron, and the coherent temperature, $T_{coh}$, below which all f-electrons are coherently screened to form a Fermi liquid. Recent experiment by Luo et al \cite{Luo2015} suggests that the protracted Kondo screening may be realized in heavy fermion compound $CeNi_{2-\delta}As_2$ where the carrier density is very low. They also found that by applying physical pressure there is a quantum phase transition from the antiferromagnetic (AFM) phase to the coherent Kondo screening state. The two phases are separated by a possible unconventional quantum
critical point. Therefore, $CeNi_{2-\delta}As_2$ presents an ideal and rare platform to investigate how the Nozi$\grave{e}$res' exhaustion affects the quantum phase transition. This experiment also poses further problems concerning the perspective of electronic structures: the origin of the dilute carrier density, the topological difference of electronic  structures on both sides of quantum phase transition, and whether chemical pressure via isovalence substitution of As by P leads to similar quantum phase transition and protracted Kondo screening. 

To answer these questions, we investigate the protracted Kondo screening by calculating the electronic structures
of the stoichiometric $CeNi_2As_2$ under compression and that of the stoichiometric $CeNi_2P_2$ at the ambient pressure. We employ the first principle density functional theory plus the dynamical mean-field theory (DFT+eDMFT) \cite{Kotliar2006,Haule2010} with continuous time quantum Monte Carlo impurity solver \cite{Werner2006,Haule2007}.
Since the $Ni$ vacancies remove the conduction electrons from crystal, our results about the Nozi$\grave{e}$res' exhaustion in the stoichiometric $CeNi_2As_2$ stay valid for $CeNi_{2-\delta}As_2$. More details about the calculations can be found in the Supplementary Information (S.I.).

\begin{figure*}
\includegraphics[width=465pt, angle=0]{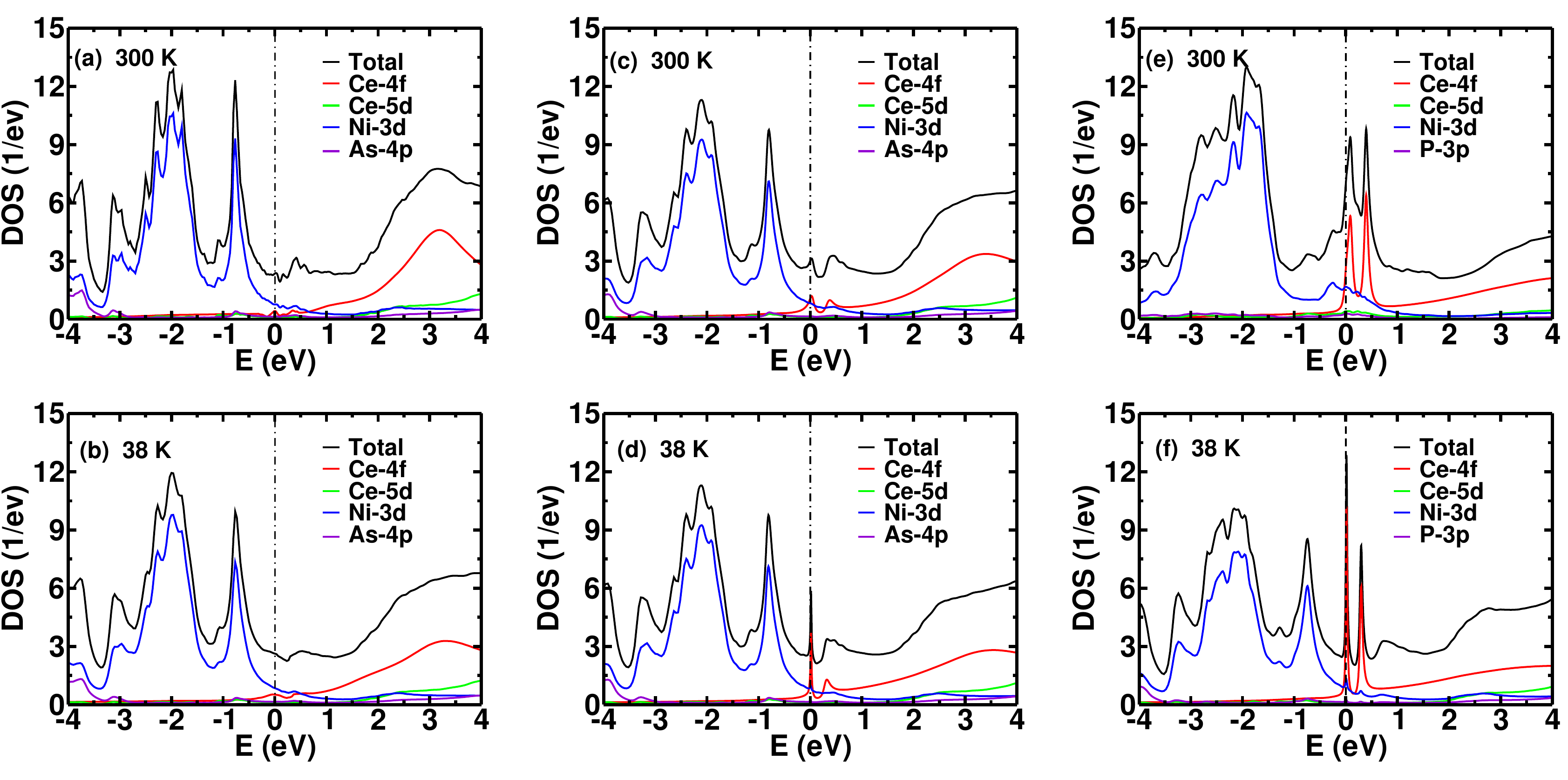}
\caption{(Color online).
The total and the partial DOS of $CeNi_2As_2$ and $CeNi_2P_2$ at 300 K (upper panels)
and at 38 K (lower panels). DOS of (a-b) $CeNi_2As_2$ at the ambient pressure,
(c-d) $CeNi_2As_2$ at 4.8 GPa, and (e-f) $CeNi_2P_2$ at the ambient pressure. The total DOS and all important partial DOS (Ce-4f, Ce-5d, Ni-3d, As-4p/P-3p) are plotted in black, red, green, blue, and violet lines respectively. The corresponding pressures of $CeNi_2As_2$ are derived from the pressure vs. volume relationship as presented in the S.I.
}
\end{figure*}

The DFT+eDMFT calculated total density of states (DOS) and all important partial DOS
(Ce-4f, Ce-5d, Ni-3d, As-4p/P-3p) of $CeNi_2As_2$
and $CeNi_2P_2$ at 300 K (upper panel)
and 38 K (lower panel) are presented in Fig.1.
In Fig.1(a-d) DOS of $CeNi_2As_2$ around the Fermi level are mainly
from the Ce-4f orbitals and the Ni-3d orbitals, although the Ce-5d orbitals and the As-4p orbitals contribute at around $\pm$4 ev.
This indicates in $CeNi_2As_2$ the conduction electrons are from Ni-3d orbitals.
In Fig.1(a-b) from 300 K to 38 K at the ambient pressure, there is no sign of enhanced hybridization
between the Ce-4f states and the conduction electrons. It proves the Ce-4f electrons of $CeNi_2As_2$
are still localized at lower temperatures and there is no Kondo resonance in $CeNi_2As_2$ at the ambient pressure (555.4 $Bohr^3$/f.u.).
However in Fig.1(c-d) when $CeNi_2As_2$ is under 4.8 GPa compression (544.2 $Bohr^3$/f.u.),
even at 300 K there is a small quasi-particle peak at the Fermi level. Further decreasing temperature down to 38 K the quasi-particle peak becomes pronounced.
The sharp quasi-particle peak comes from the hybridization between the Ni-3d states and the Ce-4f states which is enhanced by
the decreased $CeNi_2As_2$ lattice volume, in spite of the fact that the DOS of the Ni-3d orbitals around $E_F$ is still fairly small.
This is a clear manifestation of the Kondo screening.
The quasi-particle peak of the Ce-4f orbital at $E_F$
has the main quantum number $J=5/2$. The second peak of Ce-4f orbital at ~0.3 ev above $E_F$ comes
from the orbital with $J=7/2$. Our results support the experimental observation
of Luo et al \cite{Luo2015} that under compression there is a local moment to Kondo resonance phase transition in $CeNi_2As_2$.

Another interesting issue is whether chemical pressure on $CeNi_2As_2$ will lead to
the similar local moment to Kondo resonance phase transition. The chemical pressure on $CeNi_2As_2$ can be
induced via the isovalence substitution of As by P. As shown in Fig.1(e-f), there are large Kondo resonance peaks
in $CeNi_2P_2$ either at 300 K or 38 K, while the contribution of the Ni 3d-orbitals to the DOS around the Fermi energy is
still very small. This fact indicates the development of the Kondo screening states in $CeNi_2P_2$ at the
ambient pressure and the room temperature. Given the local moment state in $CeNi_2As_2$ and the Kondo screening state in
$CeNi_2P_2$, we expect a local moment to Kondo screening phase transition driven by the isovalence $As\rightarrow P$ substitution in this system. Our discovery is consistent with recent experiment observations \cite{ChenJian2017}.

\begin{figure*}
\includegraphics[width=480pt, angle=0]{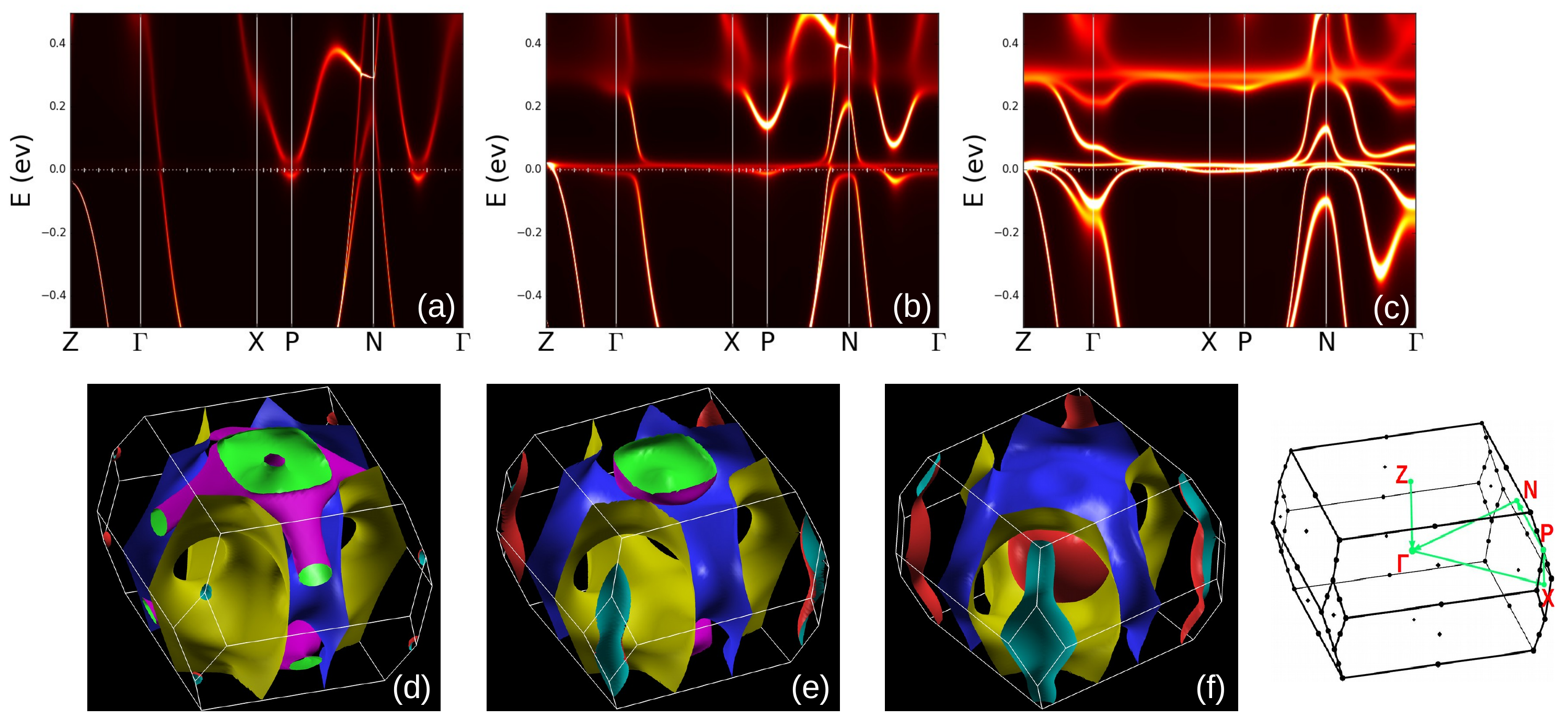}
\caption{(Color online).
The spectral function (upper panels) and the Fermi surface (lower panels) of $CeNi_2As_2$ and $CeNi_2P_2$ at 38 K.
The spectral function of (a) $CeNi_2As_2$ at the ambient pressure, (b) $CeNi_2As_2$ at 4.8 GPa, and (c) $CeNi_2P_2$ at the ambient pressure.
The Fermi surface of (d) $CeNi_2As_2$ at the ambient pressure, (e) $CeNi_2As_2$ at 4.8 GPa, and (f) $CeNi_{2}P_{e2}$ at the ambient pressure.
The k-path of spectral function in the Brillouin zone is marked by green arrows.}
\end{figure*}

It should be noticed that in Fig.1(a-b) the calculated Ni-3d DOS of $CeNi_2As_2$ is about 0.6-0.7 $ev^{-1}$ at $E_F$,
which is much smaller than the DOS at $E_F$ of conduction electrons in some other cerium based nickel pnictides like $CeNiAsO$ \cite{Luo2014} where
the Ni-3d DOS at $E_F$ is about 1.5 $ev^{-1}$. Given the Ni vacancies in the realistic $CeNi_{2-\delta}As_2(\delta\approx0.28)$ crystal, the charge carrier density in this system is even lower. According to the Nozi$\grave{e}$res' argument, the effective number of conduction electrons that participate in coherent Kondo screening is estimated by $n_{eff}=\rho_c(E_F)T_K$, where $\rho_c(E_F)$ is the DOS of conduction
electrons at $E_F$, and $T_K$ is the corresponding single impurity Kondo temperature typically much smaller than 1 ev.
Because in $CeNi_2As_2$ there is a local moment on each lattice site due to the occupied Ce-4f states,
the number of available conduction electrons is indeed small relative to the number of local moments.
Therefore, the observed coherent Kondo screening in $CeNi_2As_2$ under physical or chemical pressure are protracted.

\begin{table*}
\begin{tabular}{clclclc}
   \hline
                   & $P(GPa)$ & $\epsilon_f (ev)$ & $Im\Delta(E_F)(ev)$ & $\rho(E_F) (ev^{-1})$ & $T_K (K)$ & $T_{coh} (K)$\\
   \hline
   $CeNi_2As_2$    &  0.0   &    -2.40          &   -0.051            & 0.72                  & 3.0$\times10^{-2}$    &  9.2$\times10^{-9}$     \\
   \hline
   $CeNi_2As_2$    &  4.8   &    -2.10          &   -0.115            & 0.90                  &   86.5    &  0.1      \\
   \hline
   $CeNi_2P_2$     & 21.1   &    -2.10          &   -0.128            & 1.17                  &  147.5    &  0.366      \\
   \hline
\end{tabular}
\caption{ The pressure, the 4f electron level $\epsilon_f$, the imaginary part of hybridization function at the Fermi level $Im\Delta(E_F)$,
the Kondo temperature $T_K$, the conduction electron density at the Fermi level $\rho(E_F)$, and the coherent Kondo temperature $T_{coh}$
of $CeNi_2As_2$ and $CeNi_2P_2$. The pressure of each system
is derived from the pressure vs. volume relationship as presented in the S.I. The pressure of $CeNi_2P_2$ is the corresponding pressure of $CeNi_2As_2$ at the same volume.}
\end{table*}

To observe the low energy excitation around the Fermi level in detail, we show the momentum-resolved spectral function
$A(k,\omega)$ of $CeNi_2As_2$ and $CeNi_2P_2$ at 38 K in Fig.2(a-c). In Fig.2(a), the spectra function of
$CeNi_2As_2$ at the ambient pressure shows no sign of hybridization between the conduction
bands (mainly Ni-3d) with large dispersion and the dim flat localized Ce-4f bands at around $E_F$ and $E_F$+0.3 ev.
But in Fig.2(b), $CeNi_2As_2$ at 4.8 GPa shows strong hybridization
between the conduction bands and the Ce-4f bands. Consequently the spectra function of Ce-4f bands gain
tremendous spectra weight at $E_F$ and $E_F$+0.3 ev relative to that in Fig.2(a). 
The two enhanced Ce-4f bands in Fig.2(b) are corresponding to the $J=5/2$ and $J=7/2$ peaks in Fig.1(d) respectively.
In Fig.2(c), the spectral function of $CeNi_2P_2$ shows similar enhanced hybridization between the conduction bands
and the localized Ce-4f bands because of the chemical pressure via $As\rightarrow P$ substitution.

The Fermi surfaces of $CeNi_2As_2$ and $CeNi_2P_2$ at 38 K are also presented in Fig.2(d-f).
Unlike some other layered materials with typical two-dimensional band structures in the a-b plane 
(e.g. the structurally analogous iron pnictides $BaFe_2As_2$ \cite{PhysRevLett.101.107006, PhysRevLett.101.257003}
and the cerium based pnictides $CeNiAsO$ \cite{Luo2014} exhibiting similar quantum phase transition),
the Fermi surfaces of both $CeNi_2As_2$ and $CeNi_2P_2$ show prominent dispersion in all three directions.
The larger electron dispersion in the c-axis in $CeNi_2As_2$ and $CeNi_2P_2$ comes from the relatively shorter distance 
between the Ce-layer and the transition metal-pnictide layer. The average distance at the ambient pressure is 4.6697 Bohr in $CeNi_2As_2$ and is 4.4234 Bohr in $CeNi_2P_2$. In contrast the average distance between the Ba layer and the Fe-As layer is 6.1495 Bohr in $BaFe_2As_2$ \cite{PhysRevLett.101.257003}, and the average distance between the Ce layer and the Ni-As layer is 5.4119 Bohr in $CeNiAsO$ \cite{Luo2014}. 
In Fig.1 we found the partial DOS at the Fermi level are mainly Ce-4f and Ni-3d states. Since the Ce-4f bands show weak dispersion, it indicates in $CeNi_2As_2$ and $CeNi_2P_2$ the inter-layer hopping of Ni-3d electrons across the Ce-Ni-As/P layers is strong.
The Fermi surfaces of $CeNi_2As_2$ at the ambient pressure (Fig.2(d)) have three sheets: a large sheet (sheet 1, in green and purple) at the top and the bottom that forms hole pocket around the $N$ point, another large sheet (sheet 2, in blue and gold) surrounding the $Z-\Gamma$ line, and a tiny sheet (sheet 3, in red and cyan) barely touching the $P$ point.
Under compression to 4.8 GPa (Fig.2(e)), sheet 1 shrinks into two plates cutting the $Z-\Gamma$ line and the hole pocket at $N$ disappears, sheet 2 changes its topology to cut the $Z-\Gamma$ line as well, and sheet 3 develops into an electron pocket surrounding the $X-P$ line. 
For $CeNi_2P_2$ in Fig.2(f), sheet 1 totally disappears and sheet 2 cut the $Z-\Gamma$ line like under the physical pressure, but sheet 3 develops into two electron pockets not only surrounding the $X-P$ line but also surrounding the $\Gamma$ point. Because of enhanced hybridization between the conduction electrons and the Ce-4f electrons by increased physical/chemical pressure, more Ce-4f electrons become itinerant and the total number of electrons on the Fermi surfaces increases.

Next we estimate the two relevant energy scales, the single-impurity Kondo temperature $T_K$
and the coherent Kondo temperature $T_{coh}$ in $CeNi_2As_2$ and $CeNi_2P_2$.

When the number of
available conduction electrons $N_c$ is smaller than the number of magnetic local moments $N_f$,
$T_{coh}$ will be suppressed relative to $T_K$ according to the Nozi$\grave{e}$res' argument \cite{Nozieres1998},
\begin{equation}
 T_{coh}=\frac{N_c}{N_f}T_K=\frac{\rho(E_F)}{N_f}{T_K}^2.
\end{equation}

$T_K$ can be estimated using \cite{Gunnarsson1983,Haule2001,Pourovskii2008},
\begin{equation} 
 T_K=\sqrt{W |Im\Delta(E_F)|}e^{-\frac{\pi |\epsilon_f|}{2 N_f |Im\Delta(E_F)|}}.
\end{equation}
where $W$ is the width of conduction band below the Fermi level, $\epsilon_f$ is the average energy levels of the f-electrons,
$N_f$ is the band degeneracy of f-electrons, and $\Delta(E_F)$ is the hybridization function at the Fermi level. We choose $N_f$=6 since the Kondo peak at the Fermi level belongs to the J=5/2 Ce-4f bands and the J=7/2 peak is 0.3 ev above. The estimated Kondo temperature $T_K$ of $CeNi_2As_2$ as a function of
pressure is presented in Table. 1. At the ambient pressure the estimated $T_K$ of $CeNi_2As_2$ is roughly 0 K.
At 4.8 GPa, $T_K$ of $CeNi_2As_2$ increases to 86.5 K, which explains the enhanced Kondo peak in Fig.1(c, d). $CeNi_2P_2$ has much higher $T_K$ at 147.5 K that produces the well developed Kondo peak in Fig.1(e, f). The large $T_K$ of $CeNi_2P_2$ originates from its much smaller volume at 499.4 $Bohr^3/f.u.$. If $CeNi_2As_2$ is compressed to this volume the corresponding pressure will be 21.1 GPa.
The derived coherent temperature $T_{coh}$ is significantly lower than the Kondo temperature
$T_K$ due to the small number of effective conduction electrons $N_c=\rho(E_F)T_K$ in Kondo screening. At the ambient pressure $T_{coh}$ of
$CeNi_2As_2$ is zero since there is no Kondo screening. At 4.8 GPa $T_{coh}$ of $CeNi_2As_2$ is 0.1 K, which is much lower than the experimental result of Luo \cite{Luo2015} that the Fermi liquid temperature $T_{FL}$ is 1.0 K at 4.0 GPa. The discrepancy might come from the fact that Eq.(2) could underestimate $T_K$ since at 300 K there are obvious signs of Kondo screening in $CeNi_2As_2$ at 4.8 GPa and in $CeNi_2P_2$. Another possibility is that Nozi$\grave{e}$res' formula, Eq.(1), need a renormalization factor that depends on materials \cite{Vidhyadhiraja2000}. A more precise constrain of $T_K$ and $T_{coh}$ need to be done in the future.
For $CeNi_2P_2$, the calculated $T_{coh}$ is 0.366 K due to its larger $T_K$.
Therefore, the calculated $T_K$, $T_{coh}$ together with
the DOS in Fig.1 provide solid evidence for the development of protracted Kondo coherence in the $CeNi_2As_2$ system at low temperatures by applying
either physical pressure or chemical pressure via the isovalence $As\rightarrow P$ substitution.

By calculating the electronic structures of cerium based nickel pnictides $CeNi_2As_2$ and $CeNi_2P_2$, we find
a novel quantum phase transition from the local moment phase to the coherent Kondo screening phase in $CeNi_2As_2$ under compression. We show the coherent Kondo temperature $T_{coh}$ is much smaller than the single-impurity Kondo temperature $T_K$, and the protracted Kondo screening is due to the diluted conduction electrons on the Ni-3d orbitals. 
We further find that this transition is accompanied by topological changes of the Fermi surface. Unlike structurally analogous layered iron pnictides, the relatively stronger transfer of Ni-3d electrons across the Ce-Ni-As/P-layers 
in the z-direction makes $CeNi_2As_2$/$CeNi_2P_2$ three-dimensional materials. Our calculations also point to similar quantum phase transition and protracted Kondo screening driven by chemical pressure in agreement with the recent isovalence $As\rightarrow P$ substitution experiment. Our discoveries provide important insights in understanding the nature of the quantum phase transition and the coherent Kondo screening protraction in heavy fermion systems with dilute carrier density.

P. Z. is supported by National Science Foundation of China (NSFC) Grant No. 11604255.
J. D. is supported by NSFC Grant No. 11474082.
B. L. is supported by the National Key Research and Development Program of China (2018YFA0307600)
and NSFC Grant No. 11774282.
K. H. is supported by National Science Foundation (NSF) grant DMR-1709229.
P. Z. and J. D. want to thank J. Chen, Y. Luo, Q. Si and Z.A. Xu for extensive and valuable discussions.
This work is supported by the HPC platform of Xi'an Jiaotong university.

\bibliography{refs}
\end{document}


\title{Supplementary Materials: Protracted Kondo coherence with dilute carrier density in Cerium based nickel pnictides}

\maketitle
\newpage

{\it  DFT+eDMFT method.}
In DFT+eDMFT, the crystal problem is treated by solving the DFT equations,
whereas the strong correlations are included by solving the DMFT equations.
The outputs of DFT with self-energy correction are used to construct 
impurity levels and hybridization function for next DMFT calculation.
After the DMFT iteration is converged, new charge density and self-energy 
are derived for the next DFT iteration. Our DFT+eDMFT formalism iterates until full convergence
of charge, chemical potential, impurity levels and self-energy. 
The linear augmented plane wave WIEN2K package \cite{Blaha2001} with the generalized 
gradient approximation \cite{PBE1996} of the exchange and correlation function is used in the DFT part. 
We use a fine $12\times 12 \times 12$ k-points mesh. 
Hybridization expansion continuous time quantum Monte Carlo (CTQMC)
method is used as the impurity solver of DMFT equation.
80 million Monte Carlo steps are used at each iteration.
The local density-density form electron correlations on 4f-orbitals is parameterized by 
the Coulomb interaction U=6.0 eV and the Hund's coupling J=0.7 eV based on previous 
calculations of Cerium \cite{Lanata2013,Chakrabarti2014}. 
The fully localized limit double counting \cite{anisimov1997first} is used. 
The energy window for projector of correlated states is $\pm$10 eV around the Fermi level $E_F$.

{\it Equation of states. } 
In order to have the pressure vs. volume relationship, we have done a LDA+U with spin-orbital coupling scan of $CeNi_2As_2$ 
ranges from +2\% to -11\% of its volume $V_0$ at the ambient pressure. 
The systems are calculated using the full potential linear augmented plane wave method implemented in the Wien2k \cite{wien2k} package, within the generalized 
gradient approximation of Perdew-Burke-Ernzerhof \cite{PBE1996} for exchange and correlation. In the LDA+U calculations U-J=5.3 eV as these 
have been adopted in the DFT+eDMFT. The self-interaction correction (SIC) double counting method has been used in the calculations \cite{Anisimov1993}. 
A $12 \times 12 \times 12$ k-point mesh is used. After having converged total energy at each volume, the Murnaghan formula \cite{Murnaghan},

\vspace{0.5pt}
\begin{eqnarray*}
 E(V) & = & E_0+KV_0[\frac{1}{K'(K'-1)}(\frac{V}{V_0})^{(1-K')} + \frac{1}{K'}\frac{V}{V_0}-\frac{1}{K'-1}] \\
 P(V) & = & \frac{K}{K'} [(\frac{V_0}{V})^{K'}-1]
\end{eqnarray*}

is used to derive the pressure vs. volume relationship as shown in Figure 1 with the parameters $V_0$=555.4124 $bohr^3$, K=248.9261 GPa, K'=-4.4049, $E_0$=-32858.715733 Ry.  

\begin{center}
\begin{figure*}
\includegraphics[width=440pt, angle=0]{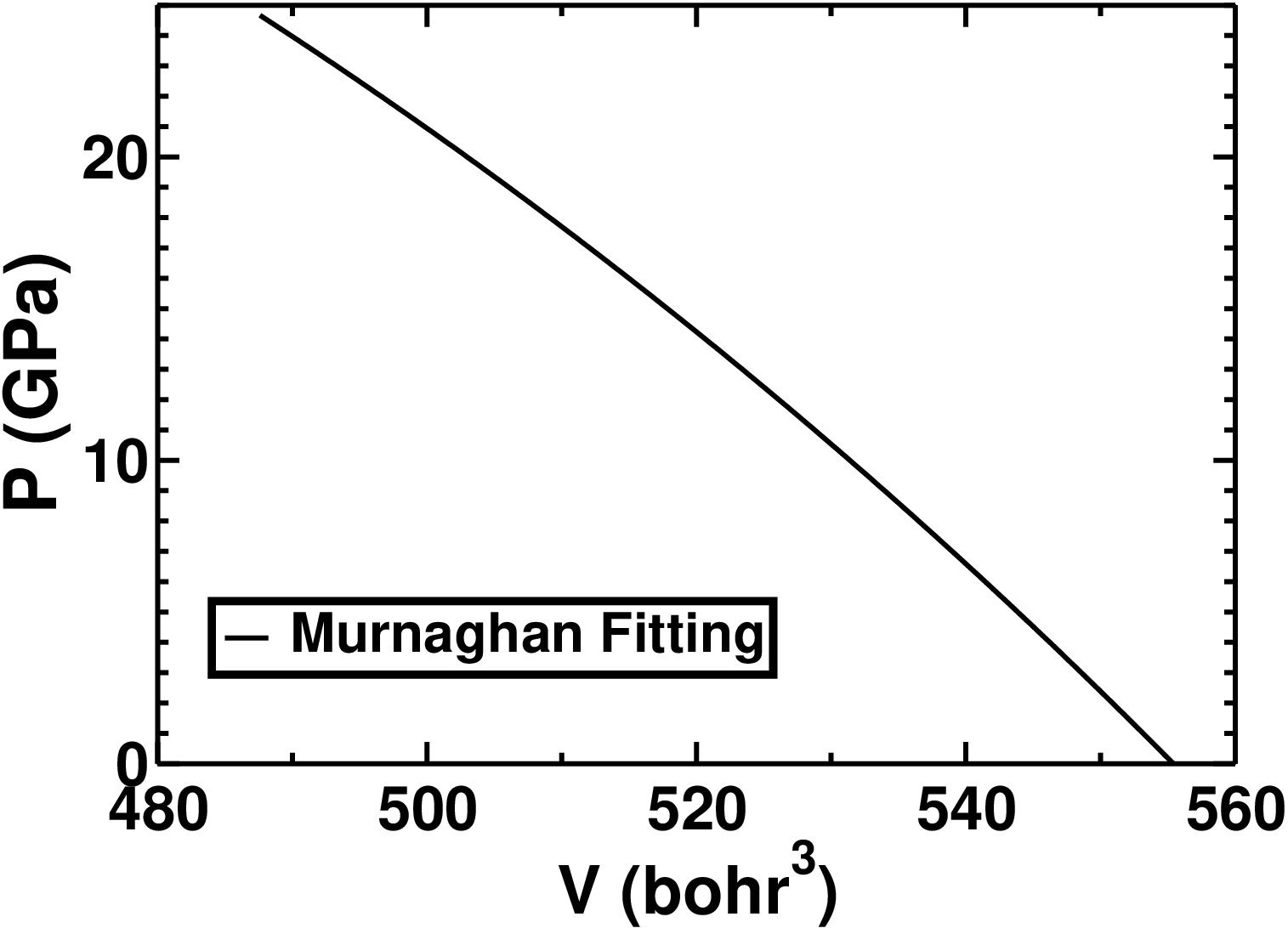}
\caption{ The pressure vs. volume relationship of $CeNi_2As_2$ as fitted using the Murnaghan formula.}
\end{figure*}
\end{center}

{\it Pressures and lattice parameters.}
Both $CeNi_{2}As_{2}$ and $CeNi_{2}P_{2}$ have body-centered tetragonal crystalline structure ($ThCr_2Si_2$-type)
of I4/mmm space group symmetry \cite{Luo2012,Bobev2009}. At the ambient pressure the lattice parameters of $CeNi_{2}As_{2}$ are
a=b=7.71127 bohr, c=18.67875 bohr, V=555.4 $bohr^3/f.u.$ with the fractional coordinate of As in unit cell $z_{As}$=0.36543 \cite{Luo2012}. After 2\% compression the lattice parameters of $CeNi_{2}As_{2}$ are a=b=7.65952 bohr, c=18.55339 bohr, V=544.2 $bohr^3/f.u.$ with the fixed lattice ratio c/a and the fractional coordinate of As $z_{As}$. The corresponding pressure is 4.8 GPa following the EOS above. At ambient pressure the 
lattice parameters of $CeNi_{2}P_{2}$ are a=b=7.47105 bohr, c=17.89287 bohr and the fractional coordinate of P is $z_{P}$=0.37354 \cite{Bobev2009}.
The volume of $CeNi_{2}P_{2}$ is 499.4 $bohr^3/f.u.$, at which a $CeNi_{2}As_{2}$ with the same volume will be under 21.1 GPa pressure. These parameters are summarized in Table 1.

\begin{table*}
\centering
\begin{tabular}{ | c | c | c | c | c | c |}
   \hline
                   &  a (bohr)  & c (bohr) & $z_{As/P}$ & V ($bohr^3/f.u.$)  &   P(GPa) \\ 
   \hline  
   $CeNi_2As_2$    &  7.71127   & 18.67875 & 0.3654     & 555.4              &    0.0   \\
   \hline
   $CeNi_2As_2$    &  7.65952   & 18.55339 & 0.3654     & 544.2              &    4.8   \\
   \hline
   $CeNi_2P_2$     &  7.47105   & 17.89287 & 0.3735     & 499.4              &   21.1   \\
   \hline   
\end{tabular}
\caption{The lattice parameters, the fractional coordinate of As/P, the volume, and the pressure of $CeNi_2As_2$ and $CeNi_2P_2$ in our calculations. The pressures of $CeNi_2As_2$ at each volume is calculated using the EOS in Fig.SI.1. The pressure of $CeNi_2P_2$ corresponds to the pressure of $CeNi_2As_2$ at the same volume.}

\end{table*}


\bibliography{refs}